\documentclass[preprint]{rsl}
\usepackage[hidelinks]{hyperref}

\title{Indoor 60 GHz Radio Channel Dataset Enabling Digital Twin Construction}
\author{Davide Scazzoli, Daniele de Santis, Francesco Linsalata, \\Fortunato Santucci, Umberto Spagnolini, Maurizio Magarini}

\begin{document}

\maketitle

%
%

\begin{abstract}
  The ambitious performance targets of modern wireless networks, including 6G and Industrial IoT (IIoT) systems, necessitate advanced hardware platforms utilizing millimeter-wave (mmWave) technology. High-frequency signals provide the bandwidth and low latency required for these systems, but rely on beamforming to overcome path loss and exploit channel sparsity. This kind of architecture provides all the specifications needed to build a SLAM (Simultaneous Localization and Mapping) system. This paper presents a dataset based on a validated, high-performance testbed integrating a Xilinx Zynq UltraScale+ RFSoC with a Sivers 60 GHz beamforming front-end. We demonstrate a novel methodology using segment-scrambled, quasi-orthogonal chirp waveforms to perform rapid exhaustive beamspace sampling. The system is integrated with Pynq Linux for real-time control and high-speed waveform upload. We present a high-density spatial dataset consisting of 350 measurement points across a 1.95 m x 3.60 m indoor grid. We exploited the system's ability to scan 63 transmit directions and construct a complete $63 \times 63$ beamspace intensity map in 200 $\mu s$. This dataset serves as a benchmark for spatial channel modeling, Digital Twins  and Integrated Sensing and Communication (ISAC) research.
\end{abstract}

\section{Introduction}

Modern industrial environments, particularly in Industry 4.0, demand autonomous systems that can operate in dynamic and often "vision-denied" conditions. In sectors such as metal manufacturing or emergency response, standard optical sensors like LiDAR often fail due to smoke, dust, or steam. Millimeter-wave (mmWave) technology offers a robust alternative, as high-frequency RF signals can penetrate smoke or fog while providing the high spatial resolution necessary for sensing~\cite{radar_lidar_low_visibility,millimap}.
The evolution toward 6G envisions a deep integration of communication and sensing (ISAC). One foundational requirement for this integration is the ability to rapidly and accurately characterize the spatial channel. Conventional Channel State Information (CSI) estimation often relies on heavy computation that is difficult to implement on real-time, resource-constrained hardware \cite{measure_indoor_ch}. To bridge this gap, Field-Programmable Gate Arrays (FPGAs) have been used to provide the deterministic, low-latency processing required for real-time beamsteering and signal synthesis~\cite{fpga_acoustic_zc,Jesus_rfsoc}.

This paper introduces a dataset for indoor radio channel mapping in both Line-of-Sight (LoS) and Non-Line-of-Sight (NLoS) conditions as well as a highly flexible prototyping methodology for mmWave sensing and communication. Our approach leverages a Radio Frequency System-on-Chip (RFSoC) platform to generate 64 semi orthogonal, constant-envelope waveforms that probe the angular domain via a phased antenna array. Current works available in the literature generally either lack an accessible dataset or probe a limited number of angles or positions~\cite{indoor_measured_vna_rotating_nopub, indoor_robottino_20pos_nopub,DISC}.

We demonstrate the system’s efficacy through a representative measurement campaign where a fixed transmitter probes a room in a grid-based acquisition. By constructing a path intensity map from signal reflections, we show that mmWave communication hardware can be repurposed for environmental mapping and self localization without the need for dedicated radar  with dedicated antenna array configuration \cite{jain2024commrad}. 
Moreover, we have acquired a dense dataset for indoor positioning, beam management and digital twin~\cite{paglierani2025digital} reconstruction which has been made available for the scientific community~\cite{dataset}.

The rest of this paper is organized as follows: Sec. ~\ref{sec:arch} introduces the system used for the dataset creation, the waveforms used and the high level design. Sec.~\ref{sec:scenario} details the scenario and main parameters used for the capture while Sec.~\ref{sec:results} shows the results obtained from the dataset and, finally, Sec.~\ref{sec:concl} concludes the paper.

\section{System Architecture}\label{sec:arch}
\begin{figure}
    \centering
    \includegraphics[width=1\linewidth]{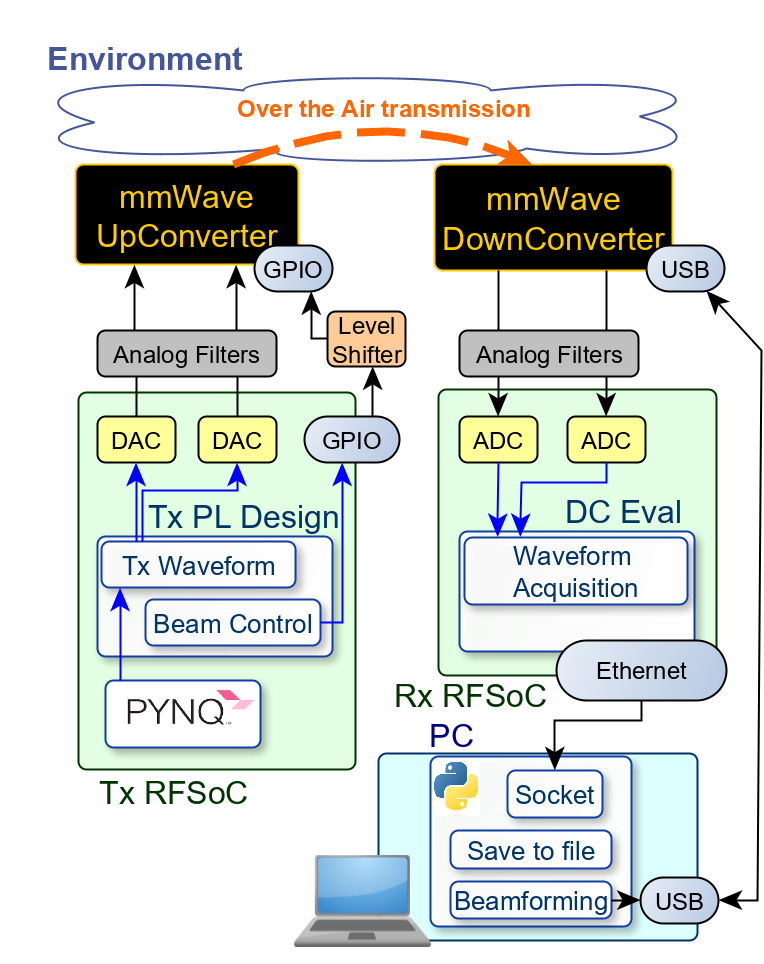}
    \caption{Main software and hardware components used in the dataset acquisition.}
    \label{fig:hwconfig}
\end{figure}

The core of the transmitter resides within the Programmable Logic (PL) of the Xilinx 1st gen. RFSoC(XCZU28DR)~\cite{ZynqEK}. 
The complete system consists of a transmitter node and a receiver node. Each node is realized with an FPGA development board and one beamformer, as shown in Fig.~\ref{fig:hwconfig}. At the transmitter side a Custom Programmable Logic (PL) is deployed with the task of transmitting the different waveforms and, at the same time, synchronously controlling the beamformer. The sampling at the receiver side has been done with the Xilinx non-MTS Evaluation Tool in conjunction with the beamformer controlled with its USB tool. The beamformers used are Sivers EVK06003~\cite{Sivers}, operating at $60.48$ GHz carrier frequency. This approach is based on several proven platforms available in the literature~\cite{RTCSA_lorenzo, Jesus_rfsoc}.

\subsection{Custom Waveform Synthesis}
To probe the angular domain with high resolution and minimal interference, we designed a custom set of 64 quasi-orthogonal signals, which lets the receiver identify the transmitted beam.
We adopted a wideband linear frequency-modulated (LFM) chirp signal, which is widely used in radar and channel sounding due to its desirable autocorrelation and spectral characteristics.

Mathematically, a baseband complex LFM chirp can be expressed as:
\begin{equation}
s(t) = \exp\left(j2\pi\left( f_0 t + \frac{f_m} {T}  t^2\right)\right), \quad 0 \leq t < T
\end{equation}
where:
\begin{itemize}
    \item $f_0 = -300 MHz$ is the chirp starting frequency
    \item $f_m = +300 MHz$ is the chirp ending frequency
    \item $f_m / T$ is the chirp rate,
    \item $B = 2\cdot f_m $ is the total bandwidth of the chirp,
    \item $T = 2.778\mu s$ is the chirp duration.
\end{itemize}

In our implementation, the base chirp spans 600 MHz, which was chosen to stay within the analog low pass filter frequency of 630 MHz for the analog filter chosen, a mini circuit VLF-630+. Correspondingly, DAC sampling rate was set to 737.28 MHz.
To create easily distinguishable signals while preserving bandwidth and power flatness, the base CHIRP is divided into $N = 64$ equal-length time segments, each of duration $T/N$. These segments are then reordered based on a pseudorandom permutation unique to each transmit beam. The function used is able to make uniform distribution over the $n!$ possible permutations with $n=64$, so the probability of each permutation is $\frac{1} {n!}$ , the algorithm is based on the pseudo-casual Mersenne Twister generator.
\\
Let the base CHIRP be segmented as:
\begin{equation}
s(t) = \sum_{n=0}^{N-1} s_n(t - n\Delta), \quad \Delta = \frac{T}{N}
\end{equation}
where $s_n(t)$ is the $n$-th time slice of the chirp, supported on $[0, \Delta]$.

For each transmit beam $k \in [0, 63]$, I define a random permutation
\\
$\pi_k: \{0,1,2,3\dots,N-1\} \rightarrow \pi_1:\{41,0,21,54,\dots,11\} ;\ \dots \space \pi_{63}:\{30,29,41,42,\dots,4\}$.
\\
The transmit signal for beam $k$ becomes:
\begin{equation}
x_k(t) = \sum_{n=0}^{N-1} s_{\pi_k(n)}(t - n\Delta)
\end{equation}

Each $x_k(t)$ therefore contains the same frequency content and energy distribution as $s(t)$ but the sub-bands are in a scrambled order that allows for effective discrimination via matched filtering.

The scrambling process preserves the essential spectral characteristics of the original chirp, as each of the 64 generated signals is a permutation of the same constituent slices. All signals occupy the exact same bandwidth. Unlike approaches based on Zadoff-Chu~\cite{fpga_acoustic_zc}, this implementation requires just one lookup table for a single chirp signal and the seed for the pseudorandom sequence.

\subsection{TX FPGA Design}

During transmission, the DACs play the scrambled chirp waveforms while also controlling the transmitter beamforming with synchronized commands.

The chirp waveform is written to memory via Direct Memory Acces (DMA), and from the DDR4 where they are loaded from Pynq Linux(running on the PS)  with a Python script. Each address of the memory blocks  corresponding to a 1024-bit data word.

Synchronization with the external beamformer is governed by a custom IP, tasked with managing the cyclic playback of the waveforms and synchronously cycling 63 directional beams plus one omnidirectional beam. Upon completion of the 64-beam cycle, it resets the BRAM read address to the initial state. Beamforming control is operated via GPIO signals: a "Beam Increment" trigger asserted after the $n$-th waveform transmission, and a "Beam Reset" signal asserted after the 64th beam.


\begin{figure}[t]
    \centering \includegraphics[width=0.94\columnwidth]{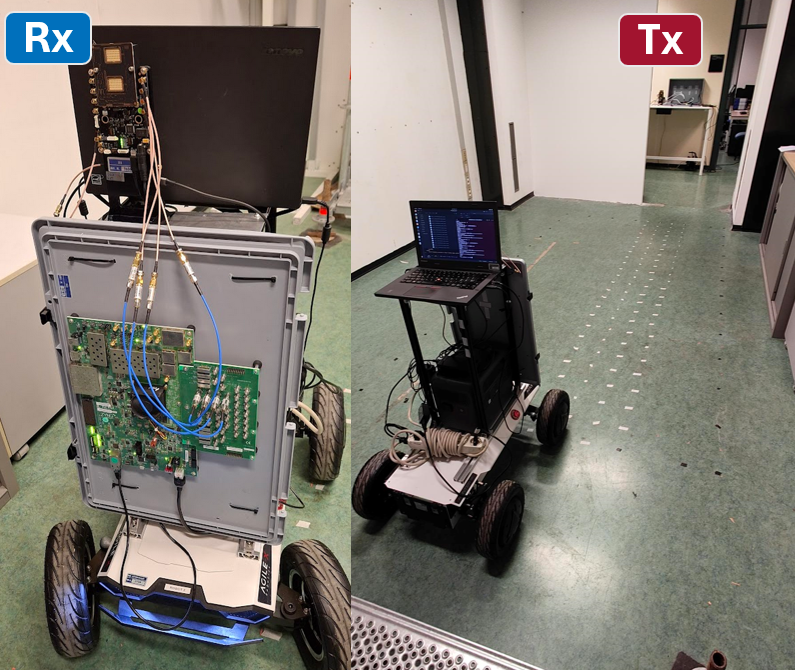}
    \caption{Photo of the mobile receiver (left) and scenario with transmitter location (right).}
    \label{fig:photo}
\end{figure}

\section{Experimental Scenario and Acquisition}
\label{sec:scenario}

To evaluate the propagation characteristics of millimeter-wave signals in a complex indoor environment, we conducted an extensive measurement campaign. The primary objective was to collect fine-grained Channel Impulse Response (CIR) data, capturing the spatial variability of the 60 GHz channel in a mixed LoS/NLoS setting. The resulting dataset has been made publicly available for the research community \cite{dataset}.

\begin{figure}[t]
    \centering \includegraphics[width=0.99\columnwidth]{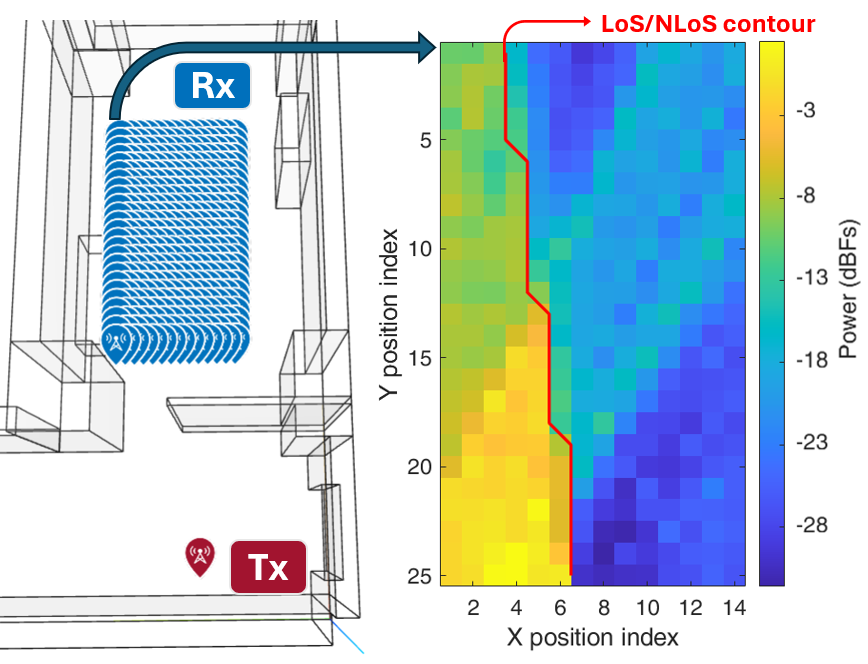}
    \caption{Spatial heatmap of the average Received Signal Strength (RSS) across the 1.95m $\times$ 3.6m grid. The contour between LoS and NLoS regions is highlighted in red.}
    \label{fig:rss_map}
\end{figure}

The measurements were performed in an indoor office environment characterized by typical structural elements such as walls, doorways, and furniture. The setup was designed to mimic a challenging industrial scenario where direct visibility is obstructed. The setup involved a fixed Transmitter (TX) and a mobile Receiver (RX), as shown in Fig.~\ref{fig:photo}. The TX node was placed in the adjacent corridor, positioned directly against the wall separating the corridor from the experimental room, as can be seen from the digital reconstruction in Fig.~\ref{fig:rss_map}. The RX was mounted on a mobile robotic platform positioned inside the room and the measurement area covered a rectangular grid of dimensions $1.95\,\text{m} \times 3.6\,\text{m}$. We acquired channel measurements at 350 distinct locations within this grid. Since the direct path between TX and RX was often obstructed by the wall, the connectivity was primarily maintained via the doorway located on the same wall or reflections, as can be seen in Fig.~\ref{fig:rss_map}. This configuration results in a received signal composed of weak penetration components and strong reflected/diffracted paths entering through the doorway, simulating a "Vision Zero" environment where radio sensing must rely on environmental bounces. For every position in the 350-point grid, we executed a double sweep of both Tx and Rx beams, resulting in directional maps such as the one reported in Fig.~\ref{fig:beam_map}. The TX, which was rotated by 18\textdegree to point directly at the entryway, scanned the environment using a codebook of 63 distinct beam sectors plus the omnidirectional one, while for each TX beam, the RX swept through 63 receive sectors. These beams scanned 21 azimuth directions, starting from -54\textdegree and ending at +54\textdegree, with three elevation angles of +18\textdegree, 0\textdegree and -18\textdegree. Due to the beamformer antenna configuration which is $2\times8$, the directivity on the elevation was much lower compared on the azimuth direction, resulting in the pattern visible in Fig.~\ref{fig:beam_map}, where leakage can be observed on the different elevations due to both low directivity and indoor multipath, which can be observed in Fig.~\ref{fig:pdp_plot}. These last are extracted by means of cross correlation between the transmitted waveforms and the acquired waveforms. Likewise, since the transmitter and receiver were not synchronized, waveform start is also identified by means of cross correlation and the code to perform this processing was made available in the dataset~\cite{dataset}.

\section{Experimental Results}
\label{sec:results}

In this section, we present the analysis of the measurement campaign conducted on the collected dataset. The post-processing focuses on the spatial distribution of the received power, the directional characteristics of the beamforming vectors, and the temporal properties of the channel.

\begin{figure}[t]
    \centering
    \includegraphics[width=0.99\columnwidth]{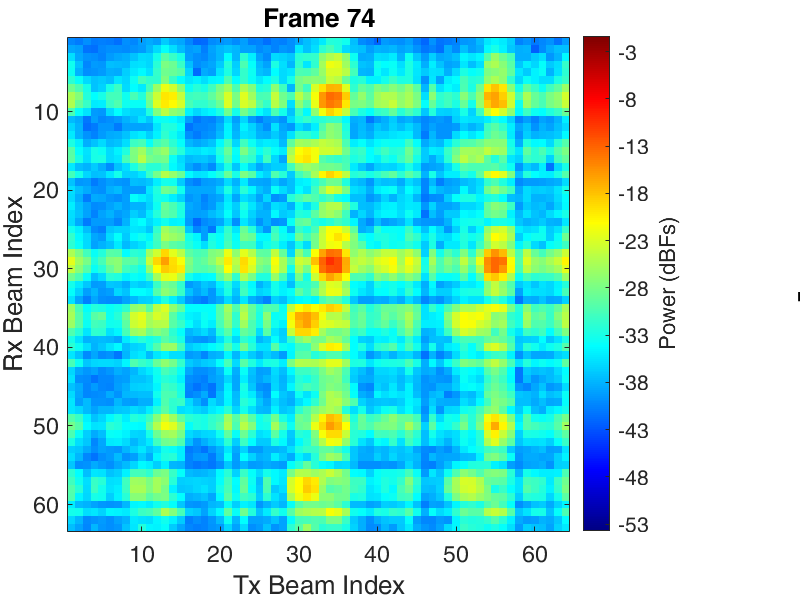}
    \caption{Map of the received powers for different Rx and Tx beams selected for position 74 corresponding to X position index of 3 and Y position index of 24.}
    \label{fig:beam_map}
\end{figure}
\begin{figure}[tpb]
    \centering
    \includegraphics[width=0.99\columnwidth]{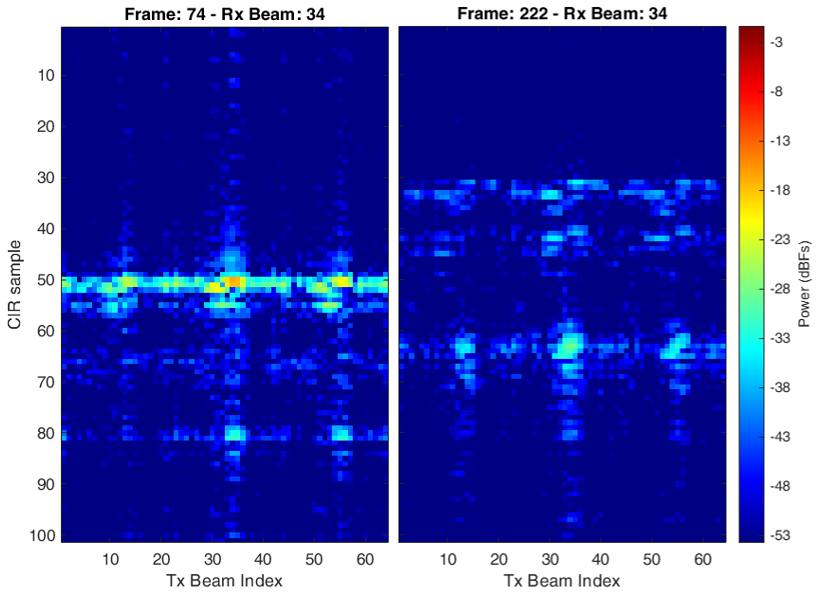}
    \caption{Power Delay Profile (PDP) of a representative LoS location (left) and NLoS location (right). Distinct multipath components are visible and their relative angle with respect to the transmitter can be identified by which transmitter beam provides the highest power.}
    \label{fig:pdp_plot}
\end{figure}

The primary metric for evaluating the link quality is the Received Signal Strength (RSS). For each of the 350 grid locations, we extracted the average RSS value across all possible TX-RX beam pairs. Fig.~\ref{fig:rss_map} illustrates the spatial heatmap of the received power. The figure shows the dataset's ability to extract environmental reflections via a simple average on the received power. A high-power "illumination zone" is clearly visible corresponding to the paths entering through the doorway. Conversely, the signal undergoes significant attenuation in the shadowed regions directly behind the concrete wall, confirming the high penetration loss typical of 60 GHz signals. However, the presence of the metallic reflectors at the entrance, visible in the photo shown in Fig.~\ref{fig:photo}, successfully extends the coverage, scattering energy into the room and ensuring that the signal remains above the noise floor even in deep NLOS positions.

To understand the geometrical consistency of the channel, we analyzed the "Best Beam Index" for each location—defined as the TX beam index that maximized the SNR. 
The analysis reveals a strong spatial correlation in beam selection. As shown in Fig.~\ref{fig:beam_map}, the majority of the grid points select beam indices that spatially align with the doorway aperture. The particular frame chosen in this figure shows both the direct LoS component, centered around Rx Beam 29 and Tx Beam 33. These beams are not specular with respect to beam 32, which corresponds to the 0\textdegree direction in both azimuth and elevation, due to the rotation angle of the transmitter which was 18\textdegree to point directly towards the entrance. Alongside the LoS, it is possible to identify a reflection from the angular map at Rx Beam 35 and Tx Beam 31, which is a reflection coming from the metal ventilation grid present on the column near the entrance of the room.

Finally, we examined the Channel Impulse Response (CIR) to characterize the multipath richness. Fig.~\ref{fig:pdp_plot} depicts the Power Delay Profile (PDP), which is a direct estimate of the CIR, for one of the receiver beams selected from the same frame shown in Fig.~\ref{fig:beam_map}. 
Due to the unsynchronized nature of the dataset, the PDPs are aligned on the average of the peaks for each Rx beam, however, across multiple Rx beams the peaks may not correspond to the same Time of Flight.

Unlike free-space scenarios, the PDP exhibits a complex structure with multiple distinct peaks. Particularly, for the NLoS case, the first arriving path is followed by stronger, delayed components corresponding to the reflections from the doorway, the metal grills, metal closet and and the far walls. This temporal diversity in reflections, together with the angular diversity, can enable self localization and environment reconstruction.

\section{Conclusions}
\label{sec:concl}

This paper presented a high-resolution indoor radio channel dataset at the $60$~GHz ISM band, designed to support research on beam management, Digital Twins, and Integrated Sensing and Communication (ISAC). The dataset was collected using a fully programmable RFSoC-based platform integrated with a commercial $60$~GHz phased-array front-end, enabling rapid and exhaustive beamspace sampling. Leveraging a custom PL design and a scrambled chirp waveform set, the system achieves a full transmitter beam scan in under $200\,\mu s$. The measurement campaign provides dense spatial coverage with 350 receiver locations over a $1.95\,\mathrm{m} \times 3.6\,\mathrm{m}$ area and a complete $63 \times 63$ Tx/Rx beamspace sweep at each point, including an omnidirectional transmit beam. Future extensions of the dataset may include multi-transmitter deployments, multi-user scenarios, and dynamic measurements with moving agents to capture doppler and temporal channel evolution. Synchronizing transmitter and receiver clocks would enable absolute Time-of-Flight estimation, unlocking finer-grained ranging and SLAM capabilities.

\section{Acknowledgments}

This work was supported by EU grant PE00000001 - program “RESTART” under the Italian National Recovery and Resilience Plan (PNRR) of NextGenerationEU

\bibliographystyle{unsrt}

\bibliography{biblio}

  

  



%
%
%
%
%
\noindent\small
Davide Scazzoli, Daniele de Santis, Francesco Linsalata, Umberto Spagnolini and Maurizio Magarini are with Politecnico di Milano, Department of Electronics, Information and Bioengineering, Via Giuseppe Ponzio 34/5, 20133 Milan, Italy\\[1ex]
Fortunato Santucci is with University of L'Aquila, Dipartimento di Ingegneria e scienze dell'informazione e matematica, Piazzale Pontieri, 1
67040 Monteluco di Roio (AQ), Italy\\
Davide Scazzoli and
Daniele de Santis have contributed equally to the paper and are co-first authors.\\[1ex]

\textbf{Corresponding author:} davide.scazzoli@polimi.it

\end{document}